\documentclass[a4paper]{article}
\usepackage{amssymb}
\usepackage{amsmath}
\usepackage{latexsym}
\usepackage{mathrsfs}
\usepackage{cancel}
\providecommand{\pacs}[1]{PACS numbers : #1}
\providecommand{\keywords}[1]{Keywords : #1}

\newcommand {\f}{\frac}

\begin{document}

\title{Dirac's Equation in $R$-Minkowski Spacetime}
\author{T.~Foughali \footnote{E-mail:
{\tt fougto\_74@yahoo.fr }}
\ and A.~Bouda\footnote{E-mail:
{\tt bouda\_a@yahoo.fr}}\\
Laboratoire de Physique Th\'eorique, Facult\'e des Sciences Exactes,\\
Universit\'e de Bejaia, 06000 Bejaia, Algeria\\}

\maketitle

\begin{abstract}
We recently constructed the $R$-Poincar\'e algebra from an appropriate deformed
Poisson brackets which reproduce the Fock coordinate transformation. We showed then
that the spacetime of this transformation is the de Sitter one. In this paper, we
derive in the $R$-Minkowski spacetime the Dirac equation and show that this is none
other than the Dirac equation in the de Sitter spacetime given by its conformally
flat metric. Furthermore, we propose a new approach for solving Dirac's equation
in the de Sitter spacetime using the Schr\"{o}dinger picture.
\end{abstract}

\pacs{03.30.+p, 98.80.Jk, 03.65.Pm}

\keywords{Fock's transformation, de Sitter spacetime, Dirac's equation.}

%###########################################################################%
%1. Introduction
%###########################################################################%

\section{Introduction}
Deformed Special Relativity (DSR) \cite{Ameliano1, Ameliano2, Mag-Smol1, Mag-Smol2} and Fock's
transformation \cite{Fock} are two distinct approaches of nonlinear relativity which are developed
for completely different motivations. In addition to the speed of light, DSR keeps invariant a minimal
length on the order of the Planck length, while Fock's transformation keeps invariant a length which
represents the universe radius.

By following the same way as in DSR \cite{Ghosh-Pal}, we recently
proposed an appropriate deformed Poisson brackets \cite{Bouda-Foughali}
\begin{align}
\{x^\mu,x^\nu\}&=0, \label{ra1}\\
\{x^\mu,p^\nu\}&=-\eta^{\mu\nu}+\frac{1}{R}\eta^{0\nu}x^\mu, \label{ra2}\\
\{p^\mu,p^\nu\}&=-\frac{1}{R}[p^\mu\eta^{0\nu}-p^\nu\eta^{\mu0}], \label{ra3}
\end{align}
from which we reproduced the Fock coordinate transformation
\begin{equation}  \label{fct}
t^{\prime }=\frac{\gamma(t-ux/c^2)}{\alpha_R},\quad x^{\prime }=\frac{
\gamma(x-ut)}{\alpha_R},\quad y^{\prime }=
                                         \frac{y}{\alpha_R},\quad z^{\prime}
                                                                           =\frac{p_z}{\alpha_R},
\end{equation}
where
\begin{equation}
{\alpha _{R}}=1+\frac{1}{R}\left[(\gamma -1)ct-\gamma \frac{ux}{c}\right],
\end{equation}
$R$ is the universe radius, $\gamma =(1-u^{2}/c^{2})^{-\frac{1}{2}}$,
$\eta^{\mu\nu} = (+1,-1,-1,-1)$ and $\mu ,\nu =0,1,2,3$.
Relations \eqref{fct} define the so-called $R$-Minkowski spacetime. From the above brackets,
we established the corresponding momentum transformation
\begin{equation}  \label{bb}
E^{\prime }={\alpha_R}{\gamma(E-up_x)},\quad p^{\prime }_x={\alpha_R}{
\gamma(p_x-uE/c^2)},\quad p^{\prime }_y={\alpha_R}{y},\quad p^{\prime }_z={
\alpha_R}{z}
\end{equation}
with which the four dimensional contraction $p_{\mu}x^{\mu}$ is an invariant,
allowing then a coherent description of plane waves.
Here $c$ and $R$ are invariant and in the limit $R\rightarrow \infty $,
the above transformations reduce to the Lorentz transformation for the coordinates
as well as for the energy-momentum vector.

The $R$-algebra, constituted by \eqref{ra1}, \eqref{ra2} and \eqref{ra3}, is completed in
\cite{Foughali-Bouda} by involving pure rotation generators,
\begin{equation}
M_i=\frac{1}{2}\epsilon_{ijk}J_{jk}  \label{Mi},
\end{equation}
and boost ones,
\begin{equation}
\tilde{N}_i=J_{0i},  \label{Ni1}
\end{equation}
where $J_{\mu \nu }=x_{\mu }p_{\nu }-x_{\nu }p_{\mu }$ stands for the angular momentum,
$i,j,...=1,2,3$ and $\epsilon _{ijk}$ is the Levi-Civita antisymmetric tensor
($\epsilon _{123}=1$). Following the Magpantay approach \cite{Magp}, we proposed a modified expression of the boost
generators
\begin{equation}
N_i = J_{0i}-\frac{1}{2R}\eta_{\mu\nu}x^\mu x^\nu p_i,          \label{Ni2}
\end{equation}
that allows to construct the first Casimir invariant of the theory.

The quantization was done by the substitution of $x^{\mu }$ and $p^{\mu }$ by the corresponding
operators and the Poisson brackets by commutators. The resulting phase space algebra of the $R$-Minkowski
spacetime
\begin{align}
[x^\mu,x^\nu]& = 0,  \label{xxs} \\
[x^\mu,p^\nu]& = -i\hbar(\eta^{\mu\nu}-\frac{1}{R}\eta^{0\nu}x^\mu),  \label{xps} \\
[p^\mu,p^\nu]& = -\frac{i\hbar}{R}(p^\mu\eta^{0\nu}-p^\nu\eta^{\mu0}),  \label{pps}
\end{align}
and the underlying $R$-Poincar\'{e} algebra
\begin{align}
[N_i,p_0]& = -i\hbar p_i+i\hbar\frac{1}{R}N_i, \\
[N_i,p_j]& = -i\hbar\delta_{ij}p_0-i\hbar\frac{1}{R} \varepsilon_{ijk}M_{k}, \\
[M_i,p_0]& = 0, \\
[M_i,p_j]& = i\hbar\varepsilon_{ijk}p_k, \\
[M_i,M_j]& = i\hbar\varepsilon_{ijk}M_k, \\
[M_i,N_j]& = i\hbar\varepsilon_{ijk}N_k, \\
[N_i,N_j]& = -i\hbar\varepsilon_{ijk}M_k,  \label{nnq}
\end{align}
allowed us to obtain the following expression for the first Casimir
\begin{equation}  \label{casi}
C = p^2_0 - {p^i}{p^i} + \frac{1}{R}({N}^i p^i+p^i{N}^i) - \frac{1}{R^2}{M}^i{M}^i ,
\end{equation}
which obviously reduces, in the limit $R\rightarrow \infty$, to the first Poincar\'{e} Casimir.

In the present work, we will focus in the $R$-Minkowski spacetime on the Dirac equation which is
already investigated in the context of the DSR \cite{GBMG, BMBGH, HSS}. Also, we aim to explore
more the correspondence, already established in \cite{Foughali-Bouda}, between de Sitter
spacetime and the $R$-Minkowski spacetime. This may offer more possibilities to handle the issue
of constructing physical observables in the de Sitter space \cite{Bala,Spra}. Furthermore,
we propose a new approach for solving the free Dirac equation in the de Sitter spacetime by
using the Schrodinger picture established in this context by Cot\u aescu \cite{Cotsp}.

The paper is organized as follows. In section 2, we establish the free Dirac equation
in the $R$-Minkowski spacetime and show that the obtained result is identically the
Dirac equation in a conformally flat de Sitter spacetime. In section 3, we construct the
Schr\"{o}dinger picture version of this equation and in section 4, we propose within this
picture a new procedure to solve the free Dirac equation in de Sitter spacetime. In
section 5, we give some concluding remarks.

%###########################################################################%
%2. Dirac equation in $R$-Minkowski spacetime
%###########################################################################%
\section{Dirac equation in $R$-Minkowski spacetime}

In \cite{Foughali-Bouda}, we proposed the following representation for the momentum
\begin{align}
p^0 & = i\hbar(\partial^0-\frac{1}{R}x^\mu \partial_\mu),  \label{repp0} \\
p^i & = i\hbar\partial^i,   \label{reppi}
\end{align}
with which the complete algebra \eqref{xxs}-\eqref{nnq} is satisfied, and therefore showed that
expression \eqref{casi} of the Casimir is exactly the Klein-Gordon operator of a conformal
flat metric of de Sitter spacetime. This result established a correspondence between
$R$-Minkowski spacetime and de Sitter spacetime.

For constructing the Dirac equation in the $R$-Minkowski spacetime, we
impose to the square of Dirac operator to reproduce partly
Klein-Gordon operator given by expression \eqref{casi} of the Casimir.
In view of this, let us express $C$ only in terms of the operators $x$ and $p$.
With the use of expressions \eqref{Mi} and \eqref{Ni2} and taking into account
commutators \eqref{xxs}, \eqref{xps} and \eqref{pps}, we can show that
\begin{align}
{M}^i{M}^i& = \epsilon^{ijk}\epsilon^{ilm}x^jp^kx^lp^m
                                 =\vec{x}^2 \vec{p}^2-(\vec{x}\cdot\vec{p})^2+i \hbar \vec{x}\cdot\vec{p},   \\
N^ip^i& = x^0\vec{p}^2-\vec{x}\cdot\vec{p}p^0-\frac{i \hbar}{R}\vec{x}\cdot\vec{p}-
                                 \frac{1}{2R}{x^0}^2\vec{p}^2+\frac{1}{2R} \vec{x}^2 \vec{p}^2,   \\
p^iN^i& = x^0\vec{p}^2-\vec{x}\cdot\vec{p}p^0-\frac{i \hbar}{R}\vec{x}\cdot\vec{p}-
                                 \frac{1}{2R}{x^0}^2\vec{p}^2+\frac{1}{2R} \vec{x}^2 \vec{p}^2 + 3i \hbar p_0.
\end{align}
The above vectors are three-dimensional ones. Substituting these relations in \eqref{casi},
we obtain
\begin{equation}
C =  p^2_0-\left(1-\frac{x^0}{R}\right)^2{\vec{p}}^2-\frac{2}{R}\vec{x}\cdot%
\vec{p}\,p^0+\frac{1}{R^2}(\vec{x}\cdot\vec{p})^2-\frac{3i \hbar}{R^2}\vec{x}\cdot%
\vec{p}+\frac{3i\hbar}{R}p_0.  \label{casi2}
\end{equation}
It is interesting to remark that by using this last expression and the fact that
\begin{equation}
\gamma^0 \gamma^i p_0\left(1-\frac{x^0}{R}\right) p_i + \gamma^i \gamma^0 \left(1-\frac{x^0}{R}\right) p_i p_0 = 0
\end{equation}
and
\begin{equation}
\frac{\gamma^i\gamma^0}{R}p_i x^j p_j +
\frac{\gamma^0\gamma^i}{R} x^jp_j p_i = -i\hbar\frac{\gamma^0\gamma^i}{R}p_i,
\end{equation}
we can show
\begin {eqnarray}
\left[\gamma^0 p_0+ \left(1-\frac{x^0}{R}\right)\gamma^i p_i + \frac{\gamma^0}{R}x^ip_i +i
\frac{3\hbar \gamma^0}{2R} +m c\right] \left[\gamma^0 p_0+ \left(1-\frac{x^0}{R}\right)
\gamma^i p_i \right.
 \hskip1mm&& \nonumber \\
\left. +\frac{\gamma^0}{R}x^ip_i+i\frac{3\hbar \gamma^0}{2R} -m c \right]
=
C -m^{2}c^{2}-i \hbar \left(1-\frac{x^0}{R}\right)\frac{\gamma^0\gamma^i}{R}p_i-
\frac{9\hbar^{2}}{4R^2}, \label{facto}
\end {eqnarray}
$\gamma^0$ and $\gamma^i$ being Dirac matrices. Equation \eqref{facto} suggests to write
the Dirac equation in $R$-Minkowski spacetime in the following form
\begin{equation}
\left[\gamma^0 p_0+\left(1-\frac{x^0}{R}\right)\gamma^i p_i+\frac{\gamma^0}{R}x^ip_i+i%
\frac{3\gamma^0}{2R} -m c\right]\Psi=0.  \label{deq}
\end{equation}
Indeed, the last two terms in \eqref{facto} that make the square of the Dirac
operator different from the Klein-Gordon operator are the manifestation of the
spin $1/2$ for the Dirac particle. In fact, in a curved spacetime, the
solution to the generally covariant Dirac equation is not a solution to the
generally covariant Klein-Gordon equation but to the generally covariant
Pauli-Schr\"{o}dinger equation describing spin $1/2$ particles in a
gravitational field \cite{Chap,Alsi}. In the square of the spinor covariant
derivative, the spinorial nature of the Dirac particle appears within the
terms involving Fock-Ivanenko coefficients. For more details, the presence of
these two additional terms is justified in Appendix A.

Using the representation given by \eqref{repp0} and \eqref{reppi}, proposed in the
$R$-Poincar\'{e} algebra context \cite{Foughali-Bouda}, we obtain the differential
form of \eqref{deq}
\begin{equation}
\left[i\left(1-\frac{x^0}{R}\right)\gamma^0 \partial_0+i(1-\frac{x^0}{R})\gamma^i
\partial_i+i\frac{3\gamma^0}{2R} -{m c/\hbar}\right]\Psi=0.  \label{decf1}
\end{equation}
This is exactly the Dirac equation in the conformally flat de Sitter
spacetime with a conformal factor $a(t)=(1-x^{0}/R)^{-1}$. Again,
a correspondence between $R$-Minkowski spacetime and de Sitter spacetime
is established.

%###########################################################################%
%3. The Schr\"{o}dinger picture
%###########################################################################%
\section{The Schr\"{o}dinger picture}

In order to solve equation \eqref{decf1}, we will use the Schr\"{o}dinger picture
developed in \cite{Cotsp}. We will give an exact solution of the Dirac equation
in the $R$-Minkowski or de Sitter spacetime, that transforms to the form found
by Shishkin \cite{Shis} and Cot\u{a}escu \cite{Cotsde} in the natural picture.

Cota\u{e}scu showed that the transformation $\Psi(x)\to \Psi_S(x)=W(x)\Psi(x)$
leading to the Schr\"{o}dinger picture is produced by the operator of time
dependent dilatations
\begin{equation}  \label{U}
W(x)=\exp\left[-\ln(\alpha(t))(\vec{x}\cdot\vec{\partial})\right]\,,
\end{equation}
with the following properties:
\begin{equation}  \label{Udag}
W(x)^{\dagger}= \alpha^3(t)\, W(x)^{-1} \,,
\end{equation}
and
\begin{equation}  \label{Wr1}
W(x)F(x)W(x)^{-1}=F\left(\frac{1}{\alpha(t)}x\right)\,,\quad
W(x)G(\vec{\partial})W(x)^{-1}=G\left(\alpha(t)\vec{\partial}\right),
\end{equation}
$F$ and $G$ being arbitrary functions. Setting $\alpha \equiv \xi^{-1}$, in the
$R$-Minkowski spacetime, we have $\xi = 1-x^{0}/R$ and then
\begin{equation}
W(x)=\exp\left[\frac{i}{\hbar}\ln(\xi)(\vec{x}\cdot\vec{p})\right] \label{osp}
\end{equation}
and
\begin{equation} \label{wpi}
W(x)G(\vec{p})W(x)^{-1}=G\left(\alpha(t)\vec{p}\right).
\end{equation}
Using the fact that $p_{0}$ commute with $\vec{x}\cdot \vec{p}$ and $[p_{0},\xi ]
=-i \hbar \xi/R$, we have
\begin{align}
W(x)p_0 W(x)^{-1} & = W(x)[p_0, W(x)^{-1}]+p_0     \nonumber \\
                  & = W(x)[p_0,\xi]\frac{\partial W(x)^{-1}}{\partial \xi}+p_0   \nonumber \\
                  & = p_0-\frac{\vec{x}\cdot\vec{p}}{R}.      \label{wp0}
\end{align}
By substituting in \eqref{deq} $\Psi(x)$ by $W^{-1}\Psi_S(x)$ and multiplying
at left the same equation by $W$, we obtain in the Schr\"{o}dinger picture the free
Dirac equation in the $R$-Minkowski spacetime
\begin{equation}
\left[\gamma^0 p_0+\gamma^i p_i-2\frac{\gamma^0}{R}\vec{x}\cdot\vec{p}+i\frac{%
3\gamma^0}{2R} -m c \right]\Psi_S=0,  \label{2b}
\end{equation}
where relations \eqref{wpi} and \eqref{wp0} have been used. With the representation given in
\eqref{repp0} and \eqref{reppi}, the differential form of the last equation is
\begin{equation}
\left[i\gamma^0\left(\left(1-\frac{x^0}{R}\right) \partial_0+\frac{x^i\partial_i}{R}+%
\frac{3}{2R}\right)+i\gamma^i\partial_i-{m c/\hbar}\right]\Psi_S=0 .  \label{8b}
\end{equation}
%
%
%###########################################################################%
%4. The solution
%###########################################################################%
\section{The solution}

Making the substitution  $\Psi_S =(1-x^{0}/R)^{3/2}\tilde{\Psi}$ in
equation \eqref{8b}, we obtain
\begin{equation}
\left[i\gamma^0\left(\left(1-\frac{x^0}{R}\right) \partial_0+\frac{x^i\partial_i}{R}\right)
+i\gamma^i\partial_i-{m c/\hbar}\right]\tilde{\Psi}=0 .  \label{8c}
\end{equation}
We put in what follows $c=\hbar=1$. Multiplying at left \eqref{8c} by $\gamma^0$
\begin{equation}
\left[i\left(1-\frac{x^0}{R}\right) \partial_0+i\left(\alpha^i+\frac{x^i}{R}\right) \partial_i
-\gamma^0{m}\right]\tilde{\Psi}=0 .  \label{8}
\end{equation}
and using the variable $\xi$, we obtain
\begin{equation}
\left[{\xi} \partial_\xi-(R\alpha^i+{x^i}) \partial_i -i\gamma^0{Rm}\right]
\tilde{\Psi}=0.  \label{10}
\end{equation}
The function $\tilde{\Psi}$ must be a bispinor
\begin{equation}
\tilde{\Psi}=
\begin{pmatrix}\varphi\\ \chi \end{pmatrix}
\end{equation}
where $\varphi$ and $\chi$ are two spinors to be determined. Using the standard
Dirac representation for $\gamma^\mu$ matrices, equation (\ref{10}) leads to
the following system
\begin{align}
\big({\xi} \partial_\xi-{x^i} \partial_i-i{Rm}\big)\varphi-R\sigma^i
\partial_i\chi&=0,  \label{11} \\
\big({\xi} \partial_\xi-{x^i} \partial_i+ i{Rm}\big)\chi
-R\sigma^i\partial_i\varphi&=0.  \label{12}
\end{align}
Multiplying at left \eqref{12} by $R\sigma^i\partial_i$ and using \eqref{11}, we get to
\begin{equation}
\left({\xi}^2 \partial^2_\xi-2\xi{x^i} \partial_i\partial_\xi+2x^i
\partial_i+x^ix^j\partial_i\partial_j-R^2\nabla^2+i{Rm} +{R^2m^2}\right)
\varphi=0,  \label{14}
\end{equation}
which we can put in the form
\begin{equation}
\left[({\xi} \partial_\xi-x^i\partial_i)^2-({\xi} \partial_\xi-x^i
\partial_i)-R^2\nabla^2+i{Rm} +{R^2m^2}\right] \varphi=0.  \label{14}
\end{equation}
In the spherical coordinate system, $(r,\theta,\phi)$, $\nabla^2$ can be separated as
\begin{equation}
\nabla^2 =\nabla^2_r+\frac{1}{r^2}\nabla^2_{(\theta,\phi)}
=\frac{1}{r^2}\frac{\partial}{\partial r} \left(r^2\frac{\partial}{\partial r}\right)-\frac{{\vec{L}}^2}{ r^2},
\end{equation}
where $\vec{L}=\vec{x}\times\vec{p}$ is the angular momentum
generator
\begin{equation}
\vec{L}^2=-\nabla^2_{(\theta,\phi)}=-\frac{1}{\sin\theta}\frac{
\partial}{\partial \theta}\sin\theta \frac{\partial}{\partial \theta}-\frac{1
}{\sin^2\theta}\frac{\partial^2}{\partial^2\phi }.
\end{equation}
Knowing that $x^i\partial_i=r\partial_r$, equation \eqref{14}
becomes
\begin {eqnarray}
\left[({\xi} \partial_\xi-r\partial_r)^2-({\xi} \partial_\xi-r\partial_r)-
\frac{R^2}{r^2}\frac{\partial}{\partial r}\left(r^2\frac{\partial}{\partial r}\right)
\right.
\hskip30mm&& \nonumber \\
\left.
+\frac{R^2}{r^2}\vec{L}^2 +iRm+{R^2m^2}\right]\varphi=0.  \label{16}
\end {eqnarray}
This equation is separable and gives a Sturm-Liouville system. So, we can
consider the following separation scheme
\begin{equation}
\varphi=U(\xi,r)\Omega(\theta,\phi),
\end{equation}
which leads to the following system
\begin{equation}
\vec{L}^2\Omega(\theta,\phi) - \lambda\Omega(\theta,\phi) = 0,
\label{17} \\
\end{equation}
\begin{eqnarray}
\left[({\xi} \partial_\xi-r\partial_r)^2-({\xi} \partial_\xi-r\partial_r)-
\frac{R^2}{r^2}\frac{\partial}{\partial r}\left(r^2\frac{\partial}{\partial r}\right)
\right.   \hskip20mm&& \nonumber \\
\left.
+i{Rm} +{R^2m^2}+\frac{R^2}{r^2}\lambda \right] U(\xi,r)&=0  \label{18},
\end{eqnarray}
where $\lambda$ is a separation constant. The solution of (\ref{17}) is
given for $\lambda=l(l+1)$ by the spherical harmonic spinors, constituted of
the usual spherical harmonics $Y_{lm}(\theta,\phi)$ and the base spinors $\chi(s_3)$,
\begin{equation}
\Omega(\theta,\phi)\equiv\Omega^l_{j,m}(\theta,\phi)=<j,m\mid l,m^{\prime };
\frac{1}{2},s_3> Y_{lm^{\prime }}(\theta,\phi)\chi(s_3),  \label{19}
\end{equation}
where $<j,m\mid l,m^{\prime };\frac{1}{2},s_3>$ are the Clebsch-Gordon
coefficients. Explicitly, these spherical harmonic spinors are given by \cite
{Grei}
\begin{align}
\Omega^l_{j,m}(\theta,\phi)&=\frac{1}{\sqrt{2j}}\begin{pmatrix}
\sqrt{j+m} \; Y_{l,m-\f{1}{2}}\\\sqrt{j-m} \; Y_{l,m+\f{1}{2}} \end{pmatrix},\quad {
\text{for}}\quad \quad j=l+\frac{1}{2}  \label{20} \\
\Omega^l_{j,m}(\theta,\phi)&=\frac{1}{\sqrt{2j+2}}\begin{pmatrix}-
\sqrt{j-m+1} \; Y_{l,m-\f{1}{2}}\\\sqrt{j+m+1} \; Y_{l,m+\f{1}{2}} \end{pmatrix}
,\quad {\text{for}}j=l-\frac{1}{2}.  \label{21}
\end{align}
Concerning equation \eqref{18}, if we use the following variable change
\begin{equation}
\eta={\xi}{r}, \hskip20mm  \quad\zeta={\xi},  \label{22}
\end{equation}
it will take the following separable form
\begin{eqnarray}
\frac{1}{R^2\zeta^2}\left(\zeta^2\partial^2_\zeta+i{Rm} +{R^2m^2}\right)
U_l(\zeta,\eta)
\hskip40mm&& \nonumber \\
= \left(\frac{\partial^2}{\partial \eta^2}+\frac{2}{ \eta}
\frac{\partial}{\partial \eta} -\frac{l(l+1)}{{\eta^2}}\right)
U_l(\zeta,\eta).  \label{25}
\end{eqnarray}
Thus, if we set $U_l(\zeta,\eta)=V(\zeta)Z_l(\eta)$ , we can obtain the following system
\begin{align}
\left(\zeta^2\partial^2_\zeta+{R^2\bar{\kappa}^2}{\zeta^2}+i{Rm} +{R^2m^2}\right)
V(\zeta)&=0,  \label{26} \\
\left[{\eta^2}\frac{\partial^2}{\partial \eta^2}+2{\eta}\frac{\partial}{%
\partial \eta}+{\bar{\kappa}^2}{\eta^2}-l(l+1)\right]Z_l(\eta)&=0,
\label{27}
\end{align}
where $\bar{\kappa}^2$ is a separation constant. Relation \eqref{26} is a Bessel's equation.
Its solution can be expressed in term of Hankel function of first kind \cite{Abra,Grad}
\begin{equation}
V(\zeta)=\zeta^{\frac{1}{2}}H_{{}_{±\nu_{-}}}(R\bar{\kappa} \zeta),
\end{equation}
where $\nu _{-}=1/2-iRm$. The general solution of Dirac's equation with $\nu _{-}$ is
a linear combination that contains terms with each of the above solutions. The other
solution with $-\nu _{-}$ is obtained with the same manner. Relation \eqref{27} is also
a Bessel's equation. Its solutions are the spherical Bessel functions:
$j_l(\bar{\kappa} \eta)=j_l(\bar{\kappa} {r}{\xi})$,
$y_l(\bar{\kappa} \eta)=y_l(\bar{\kappa} {r}{\xi})$ and
$h^{1,2}_l(\bar{\kappa} \eta)=h^{1,2}_l(\bar{\kappa} {r}{\xi})$.
For $|\bar{\kappa}|<0$, there is no solution that is bounded at infinity and
regular at the origin. For $|\bar{\kappa}|>0$, the only solution that is bounded
everywhere is
\begin{equation}
Z(\xi,r)=N_l j_l(\bar{\kappa} {r}{\xi}).
\end{equation}
where $N_{l}$ stands for a normalization constant. Thus, the physical solution
to \eqref{18} is
\begin{equation}
U_l(\xi,r)=N_l\xi^{\frac{1}{2}}H_{{}_{\nu_{-}}}(R\bar{\kappa} \xi)j_l(\bar{\kappa} {r}{
\xi}).  \label{32}
\end{equation}
Requiring that this solution be reduced in the limit $R\rightarrow \infty $ to the usual one of
Dirac's equation in the Minkowski spacetime imposes $\bar{\kappa}=p$, where $p$ is
the four-momentum \cite{Wach}. It follows that
\begin{equation}
U_l(\xi,r)=N_l\xi^{\frac{1}{2}}H_{{}_{\nu_{-}}}(Rp \xi)\,j_l(p {r}{\xi}).
\label{33}
\end{equation}

It remains to determine the second spinor $\chi$. Introducing  $X=p {r}{\xi}$
and using the property
\begin{equation}
\vec{\sigma}.\vec{\nabla} f(r)\Omega^l_{j,m}=\left[\frac{df(r)}{dr}+\frac{1+
\kappa}{r}f(r)\right](\vec{\sigma}.\hat{r})\Omega^l_{j,m},
\end{equation}
where
\begin{equation}
\kappa=\mp(j+\frac{1}{2})=\left\{
\begin{array}{r@{~}r@{~}l}
-(l+1) & \text{for} & j=l+\frac{1}{2} \\
l & \text{for} & j=l-\frac{1}{2}
\end{array}
\right.
\ \ \ \  \text{and} \ \ \ \   (\vec{\sigma}.\hat{r})\Omega^{\kappa}_{j,m}=-\Omega^{-
\kappa}_{j,m},
\end{equation}
we have
\begin{align}
R\sigma^i\partial_i\varphi&={N_l}R\sigma^i\partial_i\left[j_l(X)
\Omega^l_{j,m}(\theta,\phi)\right]\xi^{\frac{1}{2}}H_{{}_{\nu_{-}}}(Rp \xi)
\nonumber \\
&={N_l}R\left\{\left[\frac{dj_l(X)}{dr}+\frac{1+\kappa}{r}j_l(X)\right](
\vec{\sigma}.\hat{r})\Omega^l_{j,m}\right\}\xi^{\frac{1}{2}}H_{{}_{\nu_{-}}}(Rp
\xi)  \nonumber \\
&=-{N_l}R{p}\xi^{\frac{3}{2}}H_{{}_{\nu_{-}}}(Rp \xi)j_{\bar{l}}(X)\Omega^{
\bar{l}}_{j,m}, \label{rphi}
\end{align}
where we have used the properties of Bessel's functions of integer order \cite{Abra,Grad}
\begin{equation}
\frac{dj_l}{d\varrho}(\varrho)=\frac{l}{\varrho}j_l(\varrho)-j_{l+1}(\varrho),
\quad \frac{dj_l}{d\varrho}(\varrho)=j_{l-1}(\varrho)-\frac{l+1}{\varrho}j_l(\varrho),
\quad j_{-l}=(-1)^lj_l, \nonumber
\end{equation}
and we have set $\bar{l}=2j-l$. With the use of \eqref{rphi} and of the fact that
$x^{i}\partial_{i}=r\partial_{r}$,  equation \eqref{12} gives
\begin{equation}
\big({\xi} \partial_\xi-r\partial_r+ i{Rm}\big)\chi=-{N_l}R{p}\xi^{\frac{3}{2
}}H_{{}_{\nu_{-}}}(Rp \xi)j_{\bar{l}}(X)\Omega^{\bar{l}}_{j,m}.
\end{equation}
Introducing  $\tilde{\chi} = \xi^{-1/2} \chi$, the last equation takes the form
\begin{equation}
\left({\xi} \partial_\xi-r\partial_r+ i{Rm} + \frac{1}{2}\right)\tilde{\chi}=-{N_l}R{p}\xi
H_{{}_{\nu_{-}}}(Rp \xi)j_{\bar{l}}(X)\Omega^{\bar{l}}_{j,m}.
\end{equation}
Setting $z=Rp \xi$, we get to the following equation:
\begin{equation}
\left(z \partial_z -r\partial_r+ i{Rm} + \frac{1}{2}\right)\tilde{\chi} =-{N_l}z
H_{{}_{\frac{1}{2}-iRm}}(Rp \xi)j_{\bar{l}}(X)\Omega^{\bar{l}}_{j,m}.
\end{equation}
As $X=p {r}{\xi}$, it is easy to check that
$({\xi} \partial_\xi-{r} \partial_r)j_l(X)=0$.
It follows that by using the recurrence relation for Bessel's functions
\begin{equation}
z\partial_z\mathscr{C}_\nu(z)+\nu\mathscr{C}_\nu(z)=z\mathscr{C}_{\nu-1}(z),
\end{equation}
where $\mathscr{C}\equiv J,Y,H$, and the properties of Hankel's functions of first kind
\begin{equation}
{H}^{(1)}_{-\nu}(z)=e^{i\pi\nu}H^{(1)}_{\nu}(z),\quad {H}^{(2)}_{-\nu}(z)=e^{-i\pi\nu}H^{(2)}_{\nu}(z),
\end{equation}
the solution for the spinor $\chi$ can be obtained as
\begin{equation}
\chi= -{N_l}e^{-i\pi{\nu_{-}}}H^{(1)}_{{}_{\nu_{+}}}(Rp \xi)j_{\bar{l}}(p
r\xi)\Omega^{\bar{l}}_{j,m},
\end{equation}
where $\nu_{+}=1/2+iRm$. Then, the solution for Dirac's equation in its reduced
form \eqref{8c} can be written as
\begin{equation}
\tilde{\Psi}={N_l}\xi^{\frac{1}{2}}e^{- \frac{\pi}{2}Rm}\begin{pmatrix}e^{
\f{\pi}{2}Rm}H^{(1)}_{{}_{\nu_{-}}}(Rp \xi)j_l(p
{r}{\xi})\Omega^l_{j,m}(\theta,\phi)\\ i e^{-
\f{\pi}{2}Rm}H^{(1)}_{{}_{\nu_{+}}}(Rp \xi)j_{\bar{l}}(p
{r}{\xi})\Omega^{\bar{l}}_{j,m}(\theta,\phi) \end{pmatrix}.
\end{equation}
Finally, one can write the solution of the Dirac equation, relation \eqref{8b}, as
\begin{equation}
{\Psi}_{S}=\xi^{\frac{3}{2}}\tilde{\Psi}={N_l}\xi^{2}e^{- \frac{\pi}{2}Rm}
\begin{pmatrix}e^{ \f{\pi}{2}Rm}H^{(1)}_{{}_{\nu_{-}}}(Rp \xi)j_l(p
{r}{\xi})\Omega^l_{j,m}(\theta,\phi)\\ i e^{-
\f{\pi}{2}Rm}H^{(1)}_{{}_{\nu_{+}}}(Rp \xi)j_{\bar{l}}(p
{r}{\xi})\Omega^{\bar{l}}_{j,m}(\theta,\phi) \end{pmatrix}.
\end{equation}

Let us now to determine the normalization constant ${N}_l$ by using the condition
\begin{equation}
\int_\Sigma d^3x \Psi^\dagger(x)\Psi(x)=1,
\end{equation}
where $d^{3}x=r^{2}drd\Omega$ and the integration must be done over the spacelike hypersurface
$\Sigma\equiv x^{0}=cst$. For more details, see Appendix B. The standard normalization condition
of the spherical Bessel's function being \cite{Land}
\begin{equation}
\int^{\infty}_0 r^2j_l(kr)j_{l^{\prime }}(k^{\prime }r)dr=\frac{\pi}{2k^2}%
\delta(k-k^{\prime })\delta_{ll^{\prime }},
\end{equation}
for $k={p}{\xi}$, we have
\begin{equation}
\int^{\infty}_0 r^2j_l({p}{\xi}r)j_{l^{\prime }}({p}{\xi}^{\prime }r)dr=%
\frac{\pi}{2{p^2}{\xi}^3}\delta(p-p^{\prime })\delta_{ll^{\prime }},
\end{equation}
and since the spherical harmonic spinors are normalized with respect to the relation
\begin{equation}
\int (\Omega^*)^{l^{\prime }}_{j^{\prime },m^{\prime
}}\Omega^{l}_{j,m}d\Omega=\delta_{jj^{\prime }}\delta_{ll^{\prime
}}\delta_{mm^{\prime }},
\end{equation}
it follows that
\begin{eqnarray}
|N_l|^2 \frac{\pi}{2{p^2}{\xi}^3} \xi^4e^{-\pi Rm} \left[e^{-\pi
Rm}(H^{(1)}_{\nu_+})^*(Rp \xi)H^{(1)}_{\nu_+}(Rp \xi)+\right.
\hskip20mm&& \nonumber \\
\left. e^{\pi Rm}(H^{(1)}_{\nu_-})^*(Rp
\xi)H^{(1)}_{\nu_-}(Rp \xi) \right]=1.  \label{44}
\end{eqnarray}
By using the following properties of the Hankel functions of first kind
\begin{equation}
(H^{(1,2)}_{\nu_\pm})^*=H^{(2,1)}_{\nu_\mp},\quad\quad \nu_\pm=\frac{1}{2}\pm k,
\end{equation}
\begin{equation}
e^{\pm\pi k}H^{(1)}_{\nu_\mp}(z)H^{(2)}_{\nu_\pm}(z)+e^{\mp\pi
k}H^{(1)}_{\nu_\pm}(z)H^{(2)}_{\nu_\mp}(z)=\frac{4}{\pi z},
\end{equation}
equation \eqref{44} allows to deduce that
\begin{equation}
|N_l|=\frac{{R^{\frac{1}{2}}{p}^{\frac{3}{2}}}}{2}e^{\frac{1}{2}\pi Rm}.
\end{equation}
Finally, the normalized wave function reads
\begin{equation}
{\Psi}_{S}=\frac{p}{2}\sqrt{\frac{\pi R}{2\xi r}}\xi^{2}\begin{pmatrix}e^{
\f{\pi}{2}Rm}H^{(1)}_{{}_{\nu_{-}}}(Rp \xi)j_l(p
{r}{\xi})\Omega^l_{j,m}(\theta,\phi)\\ i e^{-
\f{\pi}{2}Rm}H^{(1)}_{{}_{\nu_{+}}}(Rp \xi)j_{\bar{l}}(p
{r}{\xi})\Omega^{\bar{l}}_{j,m}(\theta,\phi). \end{pmatrix}
\end{equation}

Using the operator $W$ given by \eqref{osp}, one can easily obtain the free
Dirac equation solution ${\Psi _{NP}}$ in the natural picture
\begin{equation}
{\Psi_{NP}}=W^{-1}{\Psi}_{S}=\frac{p}{2}\sqrt{\frac{\pi R}{2 r}}\xi^{2}
\begin{pmatrix}e^{ \f{\pi}{2}Rm}H^{(1)}_{{}_{\nu_{-}}}(Rp \xi)j_l(p
{r})\Omega^l_{j,m}(\theta,\phi)\\ i e^{- \f{\pi}{2}Rm}H^{(1)}_{{}_{\nu_{+}}}(Rp
\xi)j_{\bar{l}}(p {r})\Omega^{\bar{l}}_{j,m}(\theta,\phi). \end{pmatrix}
\end{equation}
If we perform a transition to the moving chart $\{\tau ,\vec{x}\}$ with
the standard flat metric $ds^{2}=c^{2}d\tau ^{2}-e^{2\omega \tau }d\vec{x}^{2}$
with $\xi =e^{-\omega \tau }$ and $\omega =1/R$, one can
easily check that ${\Psi _{NP}}$ is exactly the same solution found by
Shishkin \cite{Shis} and by Cot\u{a}escu \& al. \cite{Cotsde}.

%###########################################################################%
%5. Conclusion
%###########################################################################%
\section{Conclusion}

In the present work, we constructed the free Dirac equation in the $R$-Minkowski spacetime.
After using a certain realization of the $R$-Poincar\'{e} algebra, it turned out that the
obtained equation is exactly the Dirac equation in the conformally flat de Sitter spacetime.
This is a further proof of the correspondence between the $R$-Minkowski and the de Sitter space
such that the physics of $R$-Poincar\'{e} algebra is the same as in the de Sitter relativity.
So, this correspondence could be used to construct well-defined physical observables in the de
Sitter spacetime.

We also presented a new method for solving Dirac equation in the conformally flat patch of
de Sitter spacetime within the Schr\"{o}dinger picture. The latter was introduced by Cot\u aescu
to investigate Dirac and Klein-Gordon equations in the context of de Sitter spacetime.

%\acknowledgments

\appendix

\subsection*{Appendix A: Squared Dirac equation in curved spacetime}

\label{SDE} In a curved spacetime, Dirac's equation is given by \cite{BD}
\begin{equation}
\left( i \, \gamma^{\mu} \, D_{\mu} - \frac{m c}{\hbar} \right) \psi(x) =
0,
\end{equation}
where $D_\mu$ is the spinor covariant derivative,
\begin{equation}
D_\mu\psi=(\partial_\mu+\Omega_\mu)\psi,
\end{equation}
and
\begin{equation}
\Omega_{\mu}(x) \equiv -%
\frac{i}{4}\,\omega_{ab\mu}(x)\,\sigma^{ab} = \frac{1}{8}\,\omega_{ab\mu}(x)%
\,[\,\gamma^a,\,\gamma^b\,]
\end{equation}
\begin{equation}
\omega^{a}_{%
\hspace{.5em}b\mu} = e_{\nu}^{\hspace{.5em}a}\, \left( \partial_{\mu}\,e^{%
\hspace{.5em}\nu}_{b}  + \, e_b^{\hspace{.5em}\sigma} \,
\Gamma^{\nu}_{\sigma\mu} \right),
\end{equation}
are respectively the Fock-Ivanenko coefficients\cite{Parker} and the spin connection,
$\Gamma^{\nu}_{\sigma\mu}$ being the Christoffel symbols. It follows that
the square of Dirac's equation \cite{Chap}
\begin{equation}
\bigl( i\gamma^{\mu}D_{\mu} + m c/\hbar \bigr)\,\bigl( i\gamma^{\nu}D_{\nu}
- m c/\hbar \bigr)\psi =0\,
\end{equation}
can be rewritten in the form
\begin{equation}
\left[g^{\mu\nu}D_{\mu}D_{\nu} -\frac{1}{2}
\sigma^{\mu\nu} K_{\mu\nu} + (m c/\hbar)^2 \right]\psi=0,  \label{a2}
\end{equation}
where
\begin{equation}
K_{\mu\nu} \equiv \frac{1}{2}(D_{\nu}\,D_{\mu} -D_{\mu}\,D_{\nu})
=\partial_{\nu}\Omega_{\mu} - \partial_{\mu}\Omega_{\nu} +
[\Omega_{\nu},\Omega_{\mu}]
\end{equation}
is the spin curvature. Equation \eqref{a2} is the generally covariant extension
of the Pauli-Schr\"{o}dinger equation that describes spin $1/2$ particles in a
gravitational field \cite{Chap,Alsi}. The first term contains, in addition to the
Klein-Gordon operator, terms involving the Fock-Ivanenko coefficients $\Omega _{\mu }$
\begin{align}
g^{\mu\nu}D_{\mu}D_{\nu}\psi&=(g^{\mu\nu}\partial_{\mu}\partial_{\nu} -
g^{\mu\nu}\Gamma^{\lambda}_{\mu\nu}\partial_{\lambda})\psi +g^{\mu\nu} \left[ %
(D_\mu\Omega_{\nu}) +2 \Omega_{\nu}\partial_{\mu} \right]\psi  \nonumber \\
&=\Box_{KG}\psi+g^{\mu\nu} \left[ (D_\mu\Omega_{\nu}) +2
\Omega_{\nu}\partial_{\mu} \right]\psi.  \label{gmn}
\end{align}
In our case, the $R$-Minkowski space corresponds to the de Sitter spacetime
given by the metric
\begin{equation}
ds^{2}=\displaystyle\frac{1}{(1-x^{0}/R)^{2}}%
\left[(dx^{0})^{2}-d\vec{x}^{2}\right]=\displaystyle\frac{1}{\xi ^{2}}\left[(dx^{0})^{2}-d%
\vec{x}^{2}\right].
\end{equation}
In the chart with $x^{0}\in (-\infty ,0],$ the tetrad field
is given by $e_{\ a}^{\mu }=\xi \,\delta _{\ a}^{\mu }$ and the spin
connection by $\omega _{\mu ab}=\displaystyle{{(R\xi)^{-1} }}\left[ \eta
_{\mu b}\delta _{a}^{0}-\eta _{\mu a}\delta _{b}^{0}\right] $. Then, the covariant
derivative reads
\begin{equation}
D_{\mu }=\partial _{\mu }+\displaystyle{\frac{1}{4R\xi }}\eta
_{\mu a}\left[ \gamma ^{0},\gamma ^{a}\right],
\end{equation}
and a simple calculus gives for the sum of the terms involving the Fock-Ivanenko coefficients $\Omega$
\begin{align}
g^{\mu\nu}  \left[ (D_\mu\Omega_{\nu}) + 2\Omega_{\nu}\partial_{\mu} \right]%
&=g^{\mu\nu}  \left[ (\partial_\mu\Omega_{\nu}) -
\Gamma^{\sigma}_{\nu\mu}\Omega_{\sigma}+\Omega_{\mu}\Omega_{\nu}
+2\Omega_{\nu}\partial_{\mu} \right] \nonumber \\
&=-\frac{1}{R}\xi\gamma^0\gamma^i\partial_i-%
\frac{3}{4R^2}.       \label{gmnn}
\end{align}
On the other hand, it is known that
\begin{equation}
- \frac{1}{2} \sigma^{\mu\nu} K_{\mu\nu}= \frac{\mathcal R}{4},  \label{kmn}
\end{equation}
where $\mathcal{R}$ is the Ricci scalar and that in the de Sitter space, we have
$\mathcal{R}=12/R^2$. Taking into account equations \eqref{gmn}, \eqref{gmnn} and \eqref{kmn},
relation \eqref{a2} can be rewritten as
\begin{equation}
\left[\Box_{KG}  -\frac{1}{R}\xi\gamma^0\gamma^i\partial_i + \frac{9}{4R^2}
+ (m c/\hbar)^2 \right]\psi=0.
\end{equation}
Comparing to \eqref{facto}, it is clear that expressions of terms which make
the square of Dirac's operator different from the one of Klein-Gordon are
established.

\subsection*{Appendix B: Remark on the integration measure}

In the natural representation (NP), the normalization of the wave function,
expressed through the canonical variable $X^\mu=x^\mu/\xi$, is given by the
usual condition in special relativity
\begin{align}
<{\Psi}_{NP},{\Psi}_{NP}>
&=\int_\Sigma d^3X \bar{\Psi}_{NP}(X)\gamma^0 \Psi_{NP}(X) \nonumber \\
&=\int_\Sigma d^3X \Psi^\dagger_{NP}(X)\Psi_{NP}(X) = 1,
\end{align}
where the integration must be done over the spacelike hypersurface $\Sigma$, determined by
$X^{0}=cst$ meaning that $\xi = cst$. So, the spacelike hypersurface at a constant time in
the $R$-Minkowski spacetime is given by
\begin{equation}
d^3X|_{T=cst}=dX\wedge dY\wedge dZ\,|_{t=cst}=d^3x \xi^{-3}.
\end{equation}
Thus, the normalization condition in the $R$-Minkowski spacetime is given by
\begin{equation}
<{\Psi}_{NP},{\Psi}_{NP}>=\int_\Sigma d^3x
\xi^{-3}\Psi^\dagger_{NP}(x)\Psi_{NP}(x)=1,
\end{equation}
where $d^3x=r^2drd\Omega$. Taking into account relation (\ref{Udag}), one can
check that in the $R$-Minkowski spacetime the wave function, $\Psi\equiv\Psi_{SP}$,
in the Schr\"{o}dinger representation, is normalized with respect to the usual
condition of special relativity
\begin{align}
<{\Psi}_{SP},{\Psi}_{SP}>
& = <{\Psi}_{NP},{\Psi}_{NP}> \nonumber \\
& = \int_\Sigma d^3x
\xi^{-3}\Psi^\dagger_{NP}(x)\Psi_{NP}(x) \nonumber \\
& = \int_\Sigma d^3x
\Psi^\dagger_{SP}(x)\Psi_{SP}(x)=1.
\end{align}


\begin{thebibliography}{99}

% ref 1
\bibitem{Ameliano1}
G. Amelino Camelia, \textit{ Nature} {\bf 418}, 34 (2002),  arXiv e-print: gr-qc/0207049.

% ref 2
\bibitem{Ameliano2}
G. Amelino Camelia, \textit{ Phys. Lett.}  {\bf B510}, 255 (2001),  arXiv e-print: hep-th/0012238.

% ref 3
\bibitem{Mag-Smol1}
J. Magueijo and L. Smolin, \textit{ Phys. Rev. Lett.} {\bf 88}, 190403 (2002),  arXiv e-print: hep-th/0112090.

% ref 4
\bibitem{Mag-Smol2}
J. Magueijo and L. Smolin, \textit{ Phys. Rev. } {\bf D67}, 044017 (2003), arXiv e-print: gr-qc/0207085.

% ref 5
\bibitem{Fock}
V. Fock, \textit{ The Theory of Space, Time and Gravitation}, Pergamon Press, Oxford, London, New York, Paris (1964).

% ref 6
\bibitem{Ghosh-Pal}
S. Ghosh and P. Pal, \textit{Phys. Rev.} {\bf D75}, 105021 (2007),  arXiv e-print: hep-th/0702159.

% ref 7
\bibitem{Bouda-Foughali}
A. Bouda and T. Foughali, \textit{Mod. Phys. Lett.} \textbf{A27}, 1250036 (2012), arXiv e-print: 1204.6397.

% ref 8
\bibitem{Foughali-Bouda}
T. Foughali and A. Bouda, \textit{Can. J. Phys.}  \textbf{93}, 734 (2015),  arXiv e-print: 1605.01943.

% ref 9
\bibitem{Magp}
J. A. Magpantay, \textit{Phys. Rev.} \textbf{D84}, 024016 (2011), arXiv e-print: 1011.3888.

% ref 10
\bibitem{GBMG}
P. Gosselin, A. B\'{e}rard, H. Mohrbach and S. Ghosh, \textit{ Phys. Lett.} {\bf B660}, 267 (2008), arXiv e-print: 0709.0579.

% ref 11
\bibitem{BMBGH}
Z. Belhadi, F. Menas, A. B\'{e}rard, P. Gosselin and H. Mohrbach, \textit{ Int. J. Mod. Phys.} {\bf A27}, 1250031 (2012), arXiv e-print: 1106.1302.

% ref 12
\bibitem{HSS}
E. Harikumar, M. Sivakumar and N. Srinivas, \textit{ Mod. Phys. Lett.} {\bf A26}, 1103 (2011), arXiv e-print: 0910.5778.

% ref 13
\bibitem{Bala}
V. Balasubramanian, J. de Boer, and D. Minic, \textit{Phys.Rev.} \textbf{D65}, 123508 (2002), arXiv e-print: hep-th/0110108.

% ref 14
\bibitem{Spra}
M. Spradlin, A. Strominger and A. Volovich, arXiv e-print: hep-th/0110007.

% ref 15
\bibitem{Cotsp}
I. I. Cot\u aescu, \textit{Mod. Phys. Lett.} \textbf{A22}, 2965 (2007).

% ref 16
\bibitem{Chap} T. Chapman and O. Cerceau, \textit{Am. J. Phys.} \textbf{52}, 994 (1984).

% ref 17
\bibitem{Alsi} P. M. Alsing, J. C. Evans and K. K. Nandi, \textit{Gen. Rel. Grav.} \textbf{33}, 1459 (2001),  arXiv e-print: gr-qc/0010065.

%ref 18

\bibitem{Shis} G. V. Shishkin, \textit{Class. Quant. Grav.} \textbf{8}, 175 (1991).

%ref 19
\bibitem{Cotsde} I. I. Cot\u aescu, R. Racoceanu and C. Crucean, \textit{Mod. Phys. Lett.} \textbf{A21}, 1313 (2006), arXiv e-print: hep-th/0602077.

%ref 20
\bibitem{Grei} W. Greiner, \textit{Relativistic Quantum Mechanics - Wave
Equations}, 3rd edition (Springer 2000).

%ref 21
\bibitem{Abra} M. Abramowitz and I. A. Stegun, \textit{Handbook of
Mathematical Functions}, Dover (1964).

%ref 22
\bibitem{Grad} I. S. Gradshteyn and I. M. Ryzhik, \textit{ Tables of Integrals, Series, and Products,
Corrected and Enlarged Edition}, Academic Press, Inc, New York (1980).

%ref 23
\bibitem{Wach} A. Wachter, \textit{Relativistic Quantum Mechanics}, Springer (2011).

%ref 24
\bibitem{Land} V. Berestetski, E. Lifchitz and L. Pitayevski, \textit{Th\'{e}orie Quantique Relativiste}, Mir, Moscow (1972).

%ref 25
\bibitem{BD} N. D. Birrel and P. C. W. Davies, \textit{Quantum Fields in Curved Space}, Cambridge University Press (1982).

%ref 26
\bibitem{Parker}
L. Parker and D.J. Toms,  \textit{ Quantum Field Theory in Curved Spacetime:
Quantized Fields and Gravity}, Cambridge University Press (2009).


\end{thebibliography}
\end{document}